\documentclass[conference]{IEEEtran}
\IEEEoverridecommandlockouts
\usepackage{cite}
\usepackage{amsmath,amssymb,amsfonts}
\usepackage{algorithmicx}
\usepackage{graphicx}
\usepackage{textcomp}
\usepackage{balance}
\usepackage{amsmath}
\usepackage[utf8]{inputenc}
\usepackage{multirow}
\usepackage[hyphens]{url}
\usepackage{algorithm,algcompatible}
\usepackage[noend]{algpseudocode}

\def\BibTeX{{\rm B\kern-.05em{\sc i\kern-.025em b}\kern-.08em
 T\kern-.1667em\lower.7ex\hbox{E}\kern-.125emX}}
\begin{document}

\title{Statistical Approaches for Initial Access \\ in mmWave 5G Systems
\thanks{This work has been supported by the Research Laboratory of Huawei Milan, Italia.}
}

\author{\IEEEauthorblockN{Hossein Soleimani, Raúl Parada, Stefano Tomasin and Michele Zorzi}
\IEEEauthorblockA{\textit{Dep. of Information Engineering} \\
\textit{University of Padova, Italy}\\
\{soleimani, rparada, tomasin, zorzi\}@dei.unipd.it}
}

\maketitle

\begin{abstract}
mmWave communication systems overcome high attenuation by using multiple antennas at both the transmitter and the receiver to perform beamforming. Upon entrance of a user equipment (UE) into a cell a scanning procedure must be performed by the base station in order to find the UE, in what is known as initial access (IA) procedure. In this paper we start from the observation that UEs are more likely to enter from some directions than from others,  as they typically move along streets, while other movements are impossible due to the presence of obstacles. Moreover, users are entering with a given time statistics, for example described by  inter-arrival times. In this context we  propose scanning strategies for  IA  that take into account the entrance statistics. In particular, we propose two approaches: a memory-less random illumination (MLRI) algorithm and a statistic and memory-based illumination (SMBI) algorithm. The MLRI algorithm scans a random sector in each slot, based on the statistics of sector entrance, without memory. The SMBI algorithm instead scans sectors in a deterministic sequence selected according to the statistics of sector entrance and time of entrance, and taking into account the fact that the user has not yet been discovered (thus including memory). We assess the performance of the proposed methods in terms of average discovery time.
\end{abstract}

\begin{IEEEkeywords}
mmWave, 5G, Initial Access, Statistical, memory-less  and memory-based  algorithms.  
\end{IEEEkeywords}

\section{Introduction}

At the end of 2016, the global mobile data traffic was increasing by 7.2 exabytes per month \cite{Cisco} and an increment up to 49 exabytes monthly is expected by 2021. Unfortunately, traditional microwave cellular bands will not be suitable to accommodate this amount of data. Hence, there is a need to study alternative technologies to deliver the constant increment of data yearly. The millimeter wave (a.k.a. mmWave) band from 30 to 300 GHz is a promising space to allocate data. However, a drawback is the high attenuation in non-line-of-sight (NLOS) scenarios as the millimeter waves are blocked by common obstacles such as trees, buildings and human bodies. In a heterogeneous deployment, common macro base stations (MBS) from typical long-term evolution (LTE) networks will be complemented by small cells to provide the desired Gigabits speed rate. Opposite to LTE MBS which typically transmit in an isotropic manner, mmWave small base stations (SBS) must transmit through directional antennas to increase the signal to noise ratio (SNR), covering a range of up to 200 meters. Similarly, the user equipment (UE) will have multiple antennas and will be able to receive signals coming from the SBS only if the UE's beamformer is directed towards the SBS. When a UE enters the cell, neither the UE nor the SBS know where the other terminal is, therefore they do not know which beamformer must be used to communicate. An initial access (IA) procedure must be started, in which the SBS sequentially sends pilot packets in  various directions and in the meantime the UE scans the beamforming directions in order to receive them. The procedure is completed when the SBS beamformer points to the UE and the UE beamformer points to the SBS.

Since the cell discovery time is an issue in mmWave networks, several approaches have been proposed to minimize the time when a user is found by the SBS. Desai et al. \cite{Desai-2014-Initial} discussed these issues from the beamforming procedure point of view. Jeong et al. \cite{Jeong-random-2015} compared both omnidirectional and directional search techniques. Ferrante et al. \cite{Ferrante-2015-Mmwave} studied the effects in terms of signal to interference plus noise ratio (SINR) during the IA when the UE is in rotational motion. They compared the performance of two algorithms, exhaustive and staged: the exhaustive algorithm scans the whole 360-degree space in a sequential manner while the staged approach defines thinner beams into wider ones. Barati et al. \cite{barati-directional-2015} introduced the concept of generalized likelihood ratio test (GLRT) where base stations periodically transmit synchronization signals to establish communication. Capone et al. presented cell discovery schemes based on location information \cite{capone-context-2015}; in \cite{capone-obstacle-2015} they also introduced a learning approach to reduce the discovery time by taking into account the presence of obstacles commonly found in real scenarios (i.e., buildings). Qi and Nekovee \cite{Qi-2016-Coordinated} proposed a coordinated initial scheme for standalone mmWave networks based on the power delay profile. Giordani et al. compared different cell discovery schemes \cite{giordani-initial-2016} to find the best trade-off between delay and coverage. They studied the performance of state-of-the-art IA schemes such as exhaustive and iterative (a.k.a. staged). Furthermore, they included a context-information based scheme where global positioning system (GPS) coordinates from the closest SBS are provided by the MBS to the UE. The UE also gets its GPS information to steer the beam to the closest small base station. This approach reduces the delay but increases the energy consumption. Abbas and Zorzi \cite{Abbas-2016-Towards} studied the previously stated IA algorithms in terms of energy consumption for beamforming considering both a low power and a high power analog-to-digital converter. They also present in \cite{Abbas2016Context} an analog beamforming receiver architecture based on context information at the user. An artificial neural network (ANN) approach is presented in \cite{Cui-2016-real} to increase the positioning accuracy. There are few works based on a geo-location database  \cite{Filippini-2017-Fast}, \cite{Devoti-2017-Facing} where, by accumulating the UE position using GPS for a more efficient IA procedure. Luo et al. \cite{Luo-2017-Initial} propose a new and auxiliary transceiver for IA which operates in a narrow subband reducing the transmission length of beacon and as a consequence the beam alignment. Park et al. \cite{Park-2017-Location} estimate the user location from few mmWave access points (AP) and optimize the beamwidth required by selecting the best AP to a given user. Li et al \cite{Li-2017-Performance} analyze the two-stage beamforming approach against the exhaustive scheme resulting in an increment of the user-perceived throughput. Wei et al. \cite{Wei-2017-Exhaustive} propose a hybrid IA scheme composed by a two-stage training where the SBS trains in the first stage and a reverse training is performed by the UE in the second stage. In \cite{Jasim-2017-Fast} the Nelder Mead method is employed for fast and low complexity IA solutions. Parada and Zorzi propose a learning approach to reduce the delay in cell discovery procedure \cite{Parada-2017-Cell} by prioritizing those sectors with higher detection probability using the historical UE's location information. Li et al. \cite{Li-2017-Design} study different IA protocols, including LTE, to find the optimal design to reduce the cell discovery delay while obtaining the highest user-perceived downlink throughput. Habib et al. \cite{Habib-2017-Millimeter} propose a hybrid algorithm comparing it with both the exhaustive and the iterative state-of-the-art technique in terms of misdetection probability and discovery delay. 

In this paper we start from the observation that UEs are more likely to enter from some directions than others. This is due to the fact that users typically move along streets, while other movements are impossible due to the presence of obstacles like buildings and walls. Note also that some streets are more trafficked than others, further unbalancing the probability that users enters from specific directions. Moreover, users are entering with a given time statistics, for example described by an inter-arrival time between two users. The time statistics are also relevant for the choice of a beamforming scanning sequence by the SBS. In this context we adopt IA scanning strategies that take into account the entrance statistics. In particular, we propose two approaches: a memory-less random illumination (MLRI) algorithm and a statistic and memory-based illumination (SMBI) algorithm. The MLRI algorithm scans a random sector in each slot, based on the statistics of sector entrance, without memory. The SMBI algorithm instead scans sectors in a deterministic sequence selected according to the statistics of sector entrance and time of entrance, and taking into account the fact that the user has not yet been discovered (thus including memory). We derive the optimal beamforming scanning sequence for both approaches that minimize the average discovery time. Then we assess the performance of the proposed methods in terms of  discovery time.

The remainder of this paper is organized as follows: Section II provides the system model and describes the IA problem and the average discovery time metric. The proposed algorithms are described in Section \ref{sec:alg}. Simulation experiments and results are described in Section \ref{sec:exp}. Finally, we conclude the paper in Section \ref{sec:Conc}.
 
\section{System Model}

We consider a mmWave cellular system and focus on the problem of IA. Both base station (BS) and UEs are equipped with multiple antennas. To this end, we suppose that the (signal) space is divided into $N$ sectors that can be separately illuminated by the BS in order to discover new UEs. Sector illumination is performed by suitably choosing beamformers that are applied to the transmitted signal. In particular, the BS will be transmitting a packet known to all users, containing also the index of the illuminated sectors, and the users attempt to decode it choosing in turn a suitable beamformer to illuminate the direction from which the signal is coming. Sectors are non-overlapping and completely cover the cell. Note that the case of overlapping sectors can be easily accommodated in our work, leading only to a more complicated mathematical notation. We indicate with $p_i$ the  discrete probability density function (PDF) of the sector of entrance of the generic user, i.e., $p_i$ is the probability that user enters the cell from sector $i=1, \ldots, N$. 

Time is divided into slots and the scanning procedure by the BS is performed by exploring sector $b_k$ in slot $k$ until the user is found. We also assume that within one slot the user explores all the directions choosing multiple beamformers, so that when the BS is illuminating the sector in which the user is, the user is discovered and the IA procedure for that user is terminated. Note that we ignore the effect of noise, channel fading and interference that could prevent the detection of the IA packet by the user, thus delaying or preventing the user discovery. Moreover, we ignore false alarm events, i.e., cases in which the user erroneously detects the IA packet (when it has not been transmitted or  with the wrong index of the illuminated sector). All these features can be achieved with high probability by suitably choosing a modulation and coding format for the IA packet. We further assume that the user remains in the sector from which he entered the cell at least until it is discovered, i.e., the IA procedure is faster than the user movement within the cell. Lastly, we assume that at most one new user is present in the system, i.e., we neglect multiple entrance until the fist user has been discovered.

The time slot of entrance of the user is a random variable, with PDF $w_k$, $k=0, 1, \ldots, \infty$. We denote with $\tau$ the discovery time, i.e., the number of slots that intervene between the entrance of the user into the cell and its discovery of the BS, i.e., the end of the IA procedure. Note that $\tau$ is a random variable, depending on the exploration sequence of the sectors by the BS and the sector of entrance of the user. In particular, we look for sector exploration strategies that  minimize the average discovery time $\mathbb E[\tau]$. In particular, let $z_{k,t}$ be the probability  of  discovering  users in slot $k$, given that it entered in slot $t$, then the average discovery time is 
\begin{equation}
\bar{\tau} = \mathbb E[\tau]= \sum_{t=1}^\infty \sum_{k=1}^{\infty} (k-t) z_{k,t}
\label{avdisc}
\end{equation}

\section{IA Algorithms}\label{sec:alg}

In this section, we introduce two IA algorithms: a memory-less random illumination (MLRI) algorithm and a statistic and memory-based illumination (SMBI) algorithm. The MLRI algorithm scans a random sector at each slot, based on $p_i$ and without memory. The SMBI algorithm instead scans sectors in a deterministic sequence selected according to the statistics $p_i$ and $w_k$, and taking into account the fact that user has not yet been discovered (thus including memory). 

\subsection{Memory-less Random Illumination Algorithm}

The MLRI algorithm randomly chooses the sector to be illuminated at each slot using the PDF $q_i$, $i=1, \ldots, N$, to be properly optimized. The algorithm does not have any memory of the illuminated sectors, as each random choice is independent of the previous ones. In this case, the probability  of  discovering  the user in slot $k$ given that it entered in slot $t$ is 
\begin{equation}
z_{k,t} = \sum_{i=1}^{N} p_i (1 - q_i)^{k-t} q_i\,.
\end{equation}

The  PDF $q_i$ that minimizes the average discovery time is obtained by solving the problem 
\begin{equation}
\begin{aligned}
& \underset{\{q_i\}}{\text{min}}
& & \mathbb E[\tau] \\
& \text{subject to}
& & \sum_{i=1}^{N}q_i = 1 \\
& \text{}
& & q_i \geq 0.  
\end{aligned}
\end{equation}
We solve the problem using the Lagrangian function 
\begin{equation}
\begin{split}
f(\{q_i\}, \lambda)= & \sum_{t=1}^k w_t \sum_{k=t}^{\infty} \sum_{i=1}^{N} p_i (1 - q_i)^{k-t} q_i (k-t) +  \\ 
&+\lambda \Bigg(\sum_{i=1}^{N}q_i - 1\Bigg)\,,   
\end{split}
\end{equation}
where $\lambda$ is the Lagrange multiplier, and with a simple change of variable we have 
\begin{equation}
f(\{q_i\}, \lambda)= \sum_{k=0}^{\infty} \sum_{i=1}^{N} p_i (1 - q_i)^{k} q_i k + \lambda \Bigg(\sum_{i=1}^{N}q_i - 1\Bigg)\,,   
\end{equation}
Computing the  derivative  of the Lagrangian function  with respect to $q_i$ provides 
\begin{equation}
\frac{\partial f(\{q_i\}, \lambda)}{\partial q_i}= - \frac{p_i^2 }{q_i^2 } +\lambda  =0 ,
\end{equation}
and from the unitary-sum constraint we immediately have
\begin{equation}
q_k = p_k\,.
\end{equation}
Therefore we randomly select the sector $b_{k}$ to be illuminated in slot $k$ according to the PDF of the sector of entrance $p_i$. 

For the MLRI algorithm, the average discovery time is therefore 
\begin{equation}
\bar{\tau}_{\rm MLRI} = \sum_{i=1}^{N} p_i^2  \sum_{k=1}^{\infty} k (1 - p_i)^k = \sum_{i=1}^{N} (1-p_i)
\end{equation}

\subsection{Statistic and Memory-based Illumination Algorithm}

The MLRI algorithm does not have memory of previously explored sectors, as $b_k$ are independently generated with a slot-invariant PDF. Instead, the SMBI algorithm illuminates sector $b_{k}$ in slot $k$ based both on the statistics of users entrance ($p_i$ and $w_k$) and on the fact that the user has not yet been discovered in slot $k$. In order to determine the optimal sequence of sector illumination $b_{k}$ that minimizes the average discovery time, we resort to the maximum-a-posteriori criterion, i.e., we maximize the probability of discovering the user given that we did not find it before. 

To this end, define the set of explored sectors between slot $t$ and slot $k$ as
\begin{equation}
\mathcal S(k,t) = \{i: \exists t \leq \ell \leq k: b_\ell = i\}\,,
\end{equation}
i.e., sector $i$ is in  $\mathcal S(k,t)$ if there exists at least one slot between $t$ and $k$ in which sector $i$ has been illuminated.  Now, given that the user has not yet been discovered by slot $k$, and that the previously explored sectors are $b_1, \ldots, b_{k-1}$, the  sector to be illuminated in slot $k$ in order to minimize the average discovery time is the one maximizing the probability of finding the user in slot $k$ (given prior assumptions). Therefore, the probability of finding the user in slot $k$ illuminating sector $\ell$ is
\begin{equation}
v_k(\ell) = \sum_{t=1}^k w_t \frac{p_\ell^{(t)}}{\sum_{i=1}^N p_i^{(t)}}\,,
\end{equation}
where we set to zero the probability that the user is in sector $\ell$ if we did not find it in this sector after it was entered, i.e.,
\begin{equation}
p_i^{(t)} = \begin{cases}
0  & i \in \mathcal S(k-1,t) \\
p_i & \mbox{otherwise}
\end{cases}
\end{equation}
The  sector to be illuminated is then selected as
\begin{equation}
b_k = {\rm argmax}_{\ell} v_k(\ell)\,,
\label{bdet}
\end{equation}
and this choice  can be performed sequentially, from the first slot (for which $b_1$ is the sector having the maximum probability $p_i$), to the next slots. The resulting sequence is a deterministic, therefore as long as the statistics $p_i$ and $w_i$ do not change we will use sequence $\{b_k\}$ obtained from (\ref{bdet}) to discover all users in the system.

\section{Numerical Results}\label{sec:exp}

We now assess the performance of the proposed IA algorithms. We assume that we have $N= 17$ sectors. For comparison purposes we also consider the exhaustive-search algorithm (EA), that illuminates sectors using a periodic sequence $b_k$, where a period is a random permutation of the sector indices $1, \ldots, N$. In this case the average discovery time is 
\begin{equation}
\label{eq:exhau}
\bar{\tau}_{\rm EA} =\sum_{i=0}^{\infty}i \left [ \prod_{j=1}^{i-1} \left  ( \frac{N-j-1}{N-j}  \right ) \right]\left ( \frac{1}{N-i} \right ) \approx \frac{N}{2} 
\end{equation}

\begin{figure}[H]
    \centering
    \includegraphics[width=0.49\textwidth]{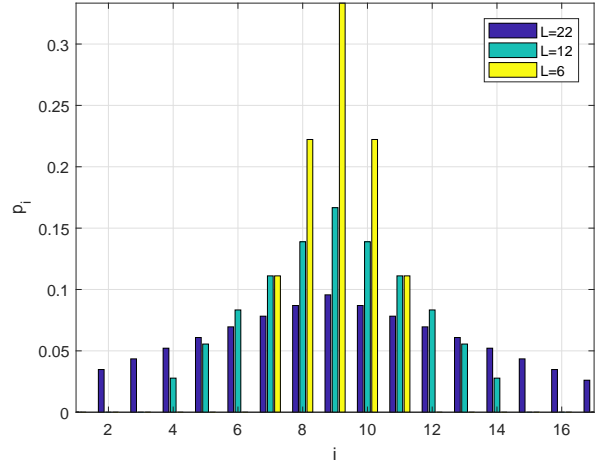}
    \caption{Sector entrance distribution $p_i$.}
    \label{fig:dist_p}
\end{figure}

The distribution $p_i$ depends on the topology of the cell, the presence of dominant paths for the users due for example to the presence of streets and obstacles like building, together with user habits. This distribution can either obtained from the topology of the cell and side information on user behavior, or estimated by the BS, based on the IA procedure of previous users. Here for demonstration purposes we consider a equilateral triangular PDF  $p_i$ with parameter $L$, i.e.,
\begin{equation}
p_{u}=\begin{cases}
\frac{4}{L^2} \left [u-[\frac{N}{2}-\frac{L}{2}] \right ] &  u \epsilon \left [\max\{\frac{N}{2}-\frac{L}{2}, 1\}, [\frac{N}{2}] \right ]    \\ 
\frac{4}{L^2} \left [ 1-[\frac{N}{2}-\frac{L}{2}] - u \right ] &  u \epsilon \left [[\frac{N}{2}], \min\{\frac{N}{2}+\frac{L}{2}, N\} \right ]  \\ 
0 & {\rm otherwise} 
\end{cases}
\end{equation}
where users never enter from sectors outside the interval $[N/2 - L/2, N/2 + L/2]$ and the probability in this interval has an equilateral triangular shape. Fig. \ref{fig:dist_p} shows some examples of the distribution $p_i$ for three values of $L$. This allows to assess the performance of the proposed algorithms when the PDF of entrance sector is more or less concentrated. For $L$ going to infinity we obtain the uniform PDF, i.e., $p_i = \frac{1}{N}$ while for $L =0$ we have that users only enter from sector $N/2$.
 
Also the the PDF $w_k$ of the user time of entrance, suitable models could be provided or $w_k$ can be estimated by observations of the BS. Here we consider an exponentially distributed inter-arrival time between user entrances, therefore
\begin{equation}
w_k =\frac{e^{-\mu k}}{1-e^{-\mu}}\,,
\end{equation}
 where $\mu$  is the average inter-arrival time. 
  
The performance of the various schemes is assessed in terms of the average discovery time (\ref{avdisc}), for which we consider both the average and the PDF. As already stated, we ignores the random effects of the channel and the false alarms and missed detection probability of user discovery.


\begin{figure}[H]
    \centering
    \includegraphics[width=0.49\textwidth]{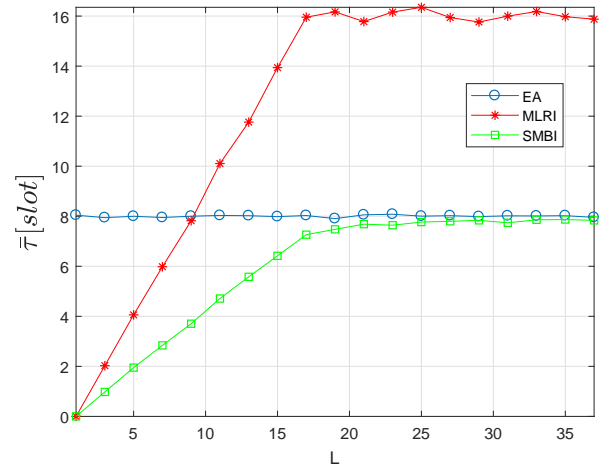}
    \caption{Average discovery time as a function of $L$ with $\mu=0.1$.}
    \label{fig:av_disc_time}
\end{figure}

\begin{figure}[H]
    \centering
    \includegraphics[width=0.49\textwidth]{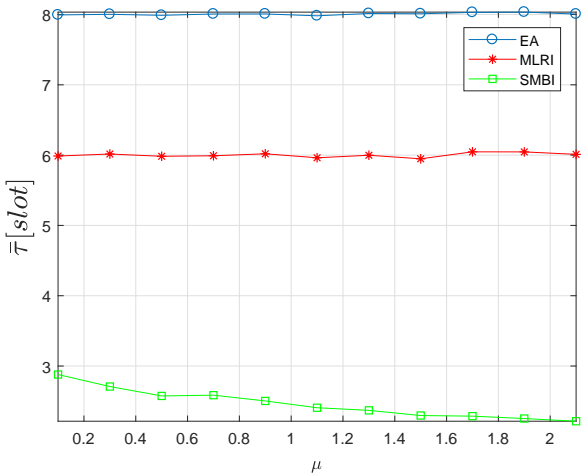}
%
    \caption{Average discovery as a function of $\mu$ for $L=8$.}
    \label{fig:av_disc_time2}
\end{figure}

In Fig. \ref{fig:av_disc_time} we first assess  the average discovery time $\bar{\tau} = \mathbb E[\tau]$ as a function of the parameter $L$ of the PDF $p_i$, for a fixed value of $\mu =0 .1$. A Monte-Carlo simulation has been run to obtain the results for the various algorithms. We can observe how the exhaustive scheme (blue-circle line) remains constant at 8 slots. The MLRI algorithm starts with a discovery time of 0 since for $L=2$ there is only one sector with probability one where the user could enter, therefore also $q_{N/2} = 1$. A similar observation holds for the SMBI algorithm that also shows a zero average discovery time. Then we observe that as $L$ increases the  average discovery time of both SMBI and MLRI increases, as the distribution $p_i$ becomes more dispersed. However, while the SMBI has the best performance achieving the minimum average discovery time, the MLRI algorithm requires even more time than the exhaustive approach for $L > 10$, as indeed with the MLRI approach it is possible that sectors are explored more than once in $N$ consecutive slots, thus being inefficient as $p_i$ tends to be uniform.
\begin{figure}[H]
    \centering
    \includegraphics[width=0.49\textwidth]{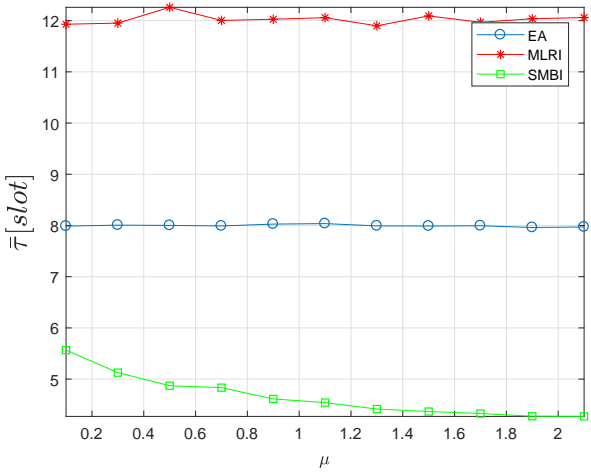}
    \caption{Average discovery as a function of  $\mu$ for $L=14$.}
    \label{fig:av_disc_time1}
\end{figure}
Fig.s \ref{fig:av_disc_time1} and \ref{fig:av_disc_time2}  show the average discovery time in the different  value of $\mu$ for a large and small  triangle window size of $L= 14$ and  $L= 8$, respectively. We observe that the SMBI method has the lowest average discovery time, thanks to the exploitation of memory. As already observed,  for small $L$  MLRI outperforms EA while for high $L$  EA  outperforms MLRI. Moreover both MLRI and EA show a constant average with respect to $\mu$  while for SMBI the average discovery time decreases with $\mu$.

\begin{figure}[H]
    \centering
    \includegraphics[width=0.49\textwidth]{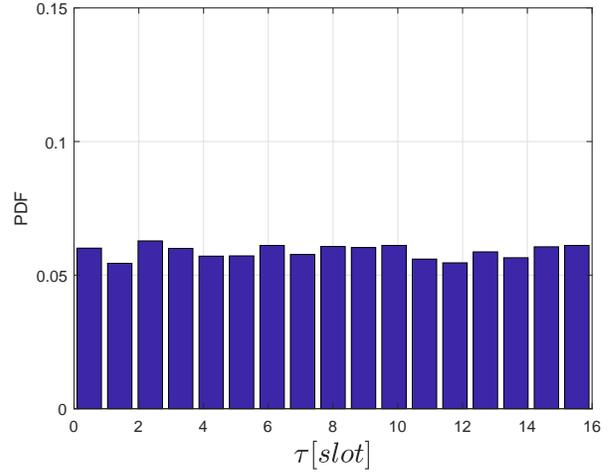}
    \caption{ PDF of discovery time for EA algorithm with $L=10$, $\mu =0.1$.}
    \label{fig:dist1}
\end{figure}

\begin{figure}[H]
    \centering
    \includegraphics[width=0.49\textwidth]{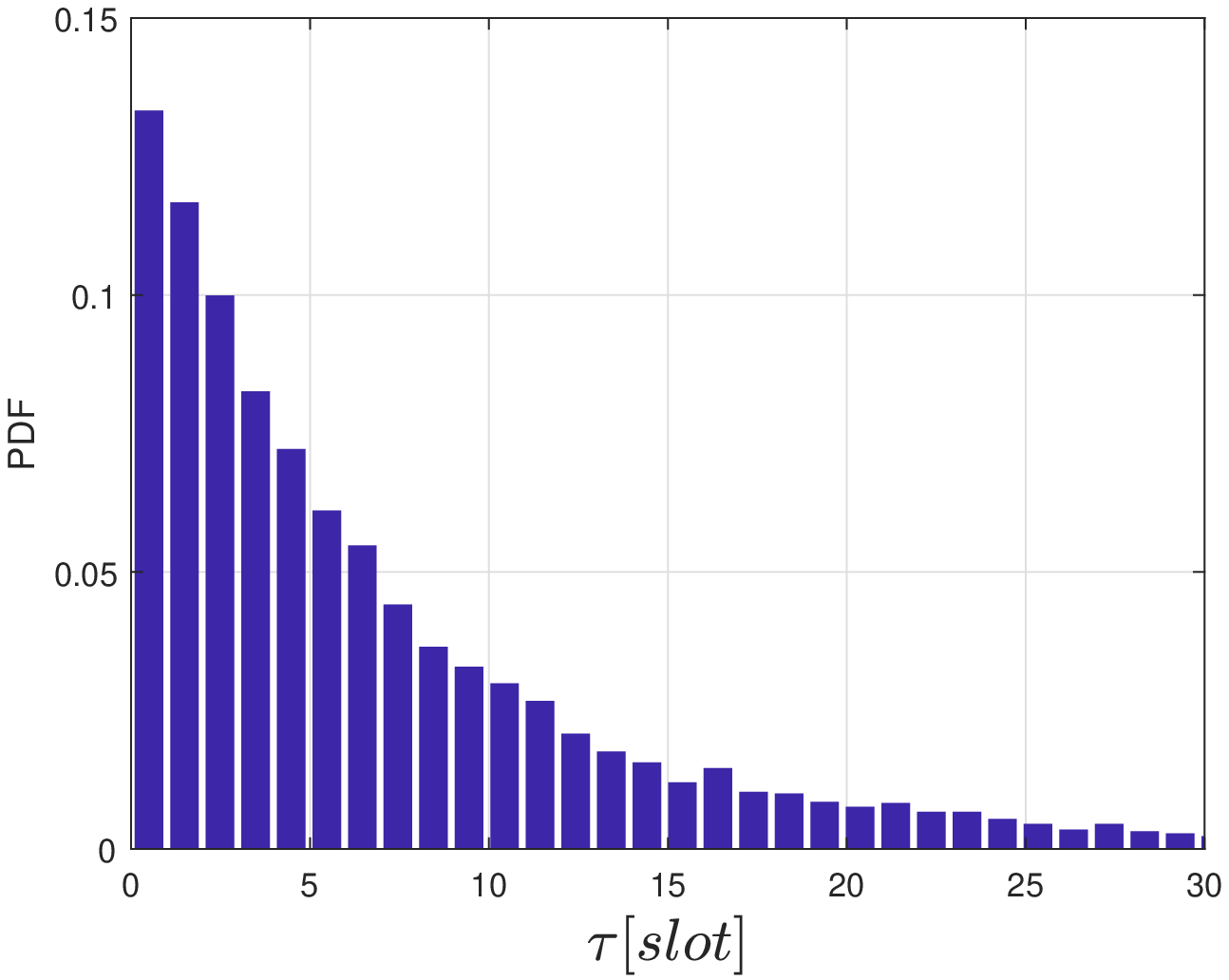}
    \caption{ PDF of discovery time for MLRI algorithm with $L=10$, $\mu =0.1$.}
    \label{fig:dist2}
\end{figure}

\begin{figure}[H]
    \centering
    \includegraphics[width=0.49\textwidth]{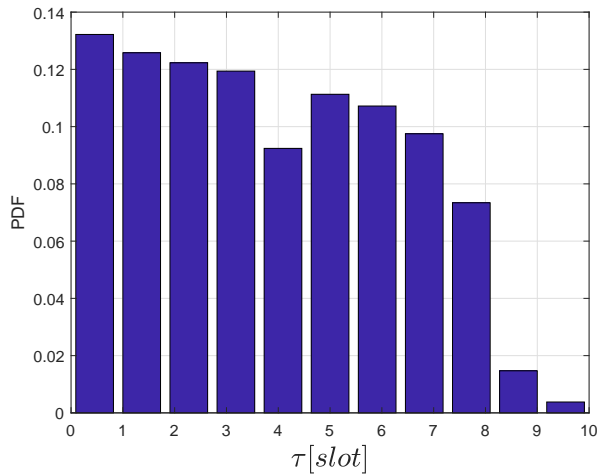}
    \caption{ PDF of discovery time for SMBI algorithm with $L=10$, $\mu =0.1$.}
    \label{fig:dist3}
\end{figure}

Lastly, Fig.s \ref{fig:dist1}-\ref{fig:dist3} show the PDF of the discovery time for the various approaches, for $\mu = 0.1$ and $L=10$. As we expect, the density function for  EA is uniform over the set $[0,N-1]$ as within $N$ slots all sectors are explored. Both MLRI and SMBI methods instead exhibit a more concentrated PDF,  in particular with the SMBI method being able to discover users within 10 slots with probability very close to 1. The MLRI method shows instead a more dispersed PDF, as expected from the results on the average discovery time.

\section{Conclusions}\label{sec:Conc}

This paper introduced two novel algorithms to reduce the average discovery time for the IA procedure for mmWave massive MIMO systems. We compared our proposed schemes with the EA and show that SMBI always outperforms EA. The simpler MLRI method instead outperforms EA only for concentrated PDF of the entrance sectors. Indeed, we are able to exploit the statistics of user entrance (both time and sector of entrance): moreover, the SMBI also exploits the memory of previously explored sectors, while the MLRI is a simpler approach based on the statistics on the sector of entrance.

\balance

\bibliographystyle{IEEEtran}
\bibliography{bibtex}

\end{document}